\documentclass[a4paper,fleqn,sort&compress]{cas-sc}
\usepackage{amsmath}

\usepackage{cite}
\usepackage[numbers]{natbib}
\usepackage{balance}
\usepackage{bm}
\def\tsc#1{\csdef{#1}{\textsc{\lowercase{#1}}\xspace}}
\tsc{WGM}
\tsc{QE}

\begin{document}

\let\WriteBookmarks\relax
\def\floatpagepagefraction{1}
\def\textpagefraction{.001}
\let\printorcid\relax

\shorttitle{Radiative metamaterials based on effective-medium theory}    

\shortauthors{H. Tan and L. Xu}  

\title [mode = title]{Radiative metamaterials based on effective-medium theory}  


% \tnotetext[1]{Title footnote text} 


\author[1]{{Haohan Tan}}

% \cormark[1]


% \ead{Email id of the first author}

% \ead[url]{URL of the first author}

\credit{Formal analysis, Writing - orginal draft}
\cormark[1]

\cortext[1]{Corresponding author.}
% \ead[url]{jphuang@fudan.edu.cn}

\ead{22110190050@m.fudan.edu.cn}
\affiliation[1]{organization={Department of Physics, State Key Laboratory of Surface Physics, and Key Laboratory of Micro and Nano Photonic Structures (MOE), Fudan University},
            % addressline={first author addressline}, 
            city={Shanghai},
%          citysep={}, % Uncomment if no comma needed betXu et al.en city and postcode
            postcode={200438}, 
            % state={first author state},
            country={China}}
         
\author[2]{{Liujun Xu}}
\affiliation[1]{organization={Graduate School of China Academy of Engineering Physics},
            % addressline={first author addressline}, 
            city={Beijing},
%          citysep={}, % Uncomment if no comma needed betXu et al.en city and postcode
            postcode={100193}, 
            % state={first author state},
            country={China}}


% \fntext[1]{Footnote text}


\begin{abstract}
  Thermal metamaterials have made significant advancements in the past few decades. However, the concept of thermal metamaterials is primarily rooted in the thermal conduction mechanism, which has consequently restricted their application scope. It is imperative to consider thermal radiation, another crucial thermal transport mechanism, particularly in high-temperature regimes, when designing thermal devices. In this review paper, we present the advancements in this area, with a specific focus on research conducted using the effective-medium theory. Additionally, we explore the potential applications of radiative thermal metamaterials and discuss prospective research directions from a microscopic perspective for future investigations.
\end{abstract}


% \begin{highlights}
% \item 
% \item 
% \item 
% \end{highlights}

% Keywords

\maketitle

% Main text
\section{Background}\label{Introduction}

The research on diffusion metamaterials dates back to 2008~\cite{XuBook23, YangPR21,
WangiScience20, HuangPhysics20, HuangESEE20, XuESEE20, Huang20, HuangESEE19, XuPRE192,
HuangPP18, JiIJMPB18, Tan20}. In that year, Huang et al. proposed thermotics. It is 
worth noting that prior to the proposition of thermotics, Huang et al.'s research focus
was not on metamaterials but on soft matter~\cite{TanJPCB20092, YangJAP07, XueCPL06,TanSM10}, 
such as colloidal particles~\cite{Tan19, TianJAP2009, 2008, XuJMR08,
FanJPCB06, LiuCTP05, HuangPRE04f, 
DongJAP041, LiuPLA04, HuangPRE03-1, HuangJAP03, HuangPRE03-2, HuangPLA02}, colloidal crystal~\cite{ZhangCPB10,FanJPCC2009, 
HanPLA2009, ZhangCPL2009, ZhangAPL08, HuangJRCC08, HuangAPL05a},
and colloidal ferrofluids~\cite{GaoPP10, FanJAP2008, WangCPL06, HuangPRE05b, HuangPRE04a}),
electrorheological fluid~\cite{QiuCTP15, Li2012sm, BaoJPCM10, WuEPJAP09, TanJPCB20091, XiaoJPCB2008,
FangCPL07, TianCPL06, CaoJPCB06, ShenCPL06, TianPRE06, HuangCP04,LiuIJMPB11}, 
and electrorheological solid~\cite{HuangPRE04b, HuangPLA04, HuangCPL04}, 
magnetorheological fluid~\cite{Li2013EPJP, HuangJPCM04, HuangPRE04c, HuangJCP04}. 
They also ventured into the research of water, including the movement of water molecules~\cite{MengCPB18, 
Tan14, Tan17, QiuJP.Phys.Chem.B15, QiuTEPJ-AP14, Meng2013pre, Wang2013ACM, 
Meng2013mp, Wang2012cpb, WangJPCB11, MengJPCB11, TianEPL07} and phase transition~\cite{Tan3, Tan7, Tan18},
electromagnetism, including electrics~\cite{ 
chen2013prl, FanCTP10, ZhaoJAP2009, GaoAPL2008, ZhuJAP08, WangJPCB06,
DongEPJB05, HuangPRE05a, HuangJPCB04, DongJAP042, HuangCTP03,
HuangCTP02}, nonlinear response~\cite{Tan8, XuPLA06, HuangPLA05, HuangJPCB05, HuangPRE04e, 
HuangPRE04g, 
HuangPRE01, HuangCTP01-1},
dispersion~\cite{ KoEPJE04, KoJPCM04, GaoPRE03}, electrorotation~\cite{TianPRE07, HuangJPCM02, HuangPRE02},
transformation electrics~\cite{QiuCTP14, GaoJAP2009}, ferrofluids~\cite{Fan2013cpb, JianJRCB08, GaoJPCC07,
HuangJMMM05,Fan2013fop, FanJPDAP11, GaoPRL10}, and other fields. Furthermore, they were also interested in sociology~\cite{JiPA18,
XinPA17, Tan5, Tan12, QiuCPB14, QiuPLA14, QiuFP14, liang2013pre, SongARCS12, ZhaoPNAS11,
WangPNAS09, GuCTP2009, ChenJPA07, DaiiScience21}, econophysics, including the stock market~\cite{Tan9, QiuPR15, QiuPLoS141, QiuPLoS142, QiuPLoS13, 
QiuEPJB13, Liang2013fopw, wei2013AnAM, wei2013plos, li2012epl, ZhouPA2009, YePA081}, and other aspects~\cite{XinFP17,Tan6, 
Tan13, QiuPLA15, QiuSpringer15, QiuJSM14},
statistics~\cite{Liu2013ASP, HuangPRE04d}, condensed matter physics~\cite{WangOL10},
and energy engineering~\cite{ShenPRL16, QiuEur.Phys.J.Appl.Phys.15}. Perhaps it is 
the study of optics, including optical materials~\cite{ huang2013cp, JianJPCC2009,
WangOL2009, WangOL2008, FanAPL06, HuangOL05, XiaoPRB05, HuangJAP05, PanPB01, HuangCTP01-2} 
and nonlinear response~\cite{HuangNY07, WangAPL07, HuangPR06a,
HuangJAP06, 
HuangJOSAB05, HuangAPL05b, HuangEL04, HuangAPL04,
GaoPRB04, GaoEPJB03, HuangSSC00},
and acoustics~\cite{ZhaoFOP12,
SuFOP11, LiuCTP10, LiuEPJAP2009}  that paved the way for the eventual birth of thermotics.

Since then, the research in the field of diffusion metamaterials has made significant progress~\cite{PJ-rmp,PJ-click,PJ-research,PJ-am,DaiPR23, XuPANS23, ZhangNRP23, XuNSR23, XuPRL222}. Researchers have designed numerous devices with novel functions, such as thermal cloaks~\cite{YaoISci22, XuIJHMT21, XuPRE20, XuEPL202, YangJAP19, QiuIJHT14, QiuEPL13, LiJAP10, FanAPL2008, Tan21, Tan26, Tan28, Tan25}, concentrators~\cite{Tan39, ZhuangSCPMA22}, rotators~\cite{YangPRAP20}, sensors~\cite{JinIJHMT21, WangEPL21, XuEPL201, JinIJHMT20, Tan24}, illusions~\cite{WangPRAP20, XuEPJB191, YangESEE19, QiuAIP Adv.15, TianIJHMT21, Tan49}, transparency~\cite{LiuJAP211, YangJAP20, XuPRA19a, WangJAP18, Tan22, Tan23}, and expanders~\cite{HuangPB17, Tan29}. The development of metamaterials in the thermal field began with research on thermal conduction~\cite{Tan36, Tan37, Tan38, Tan43, JiCTP18, HuangAMT22, XuEPJB18, ShenAPL16, WangPRE20, WangJAP17, YangAPL17}. However, thermal radiation~\cite{Tan31, Tan32, Tan33, Tan34, Tan35, Tan45, Tan46, Tan47, Tan48, ZhangCPL23, XuPRAP20, XuPRAP19, ShangAPL18}, which is an important mode of thermal transport alongside conduction and convection~\cite{DaiJAP18, Tan30, XuEPL211}, cannot be ignored in many cases. Finding ways to realize similar functions while considering this mechanism is an unavoidable problem. However, the nonlinearity of Stefan's law has posed significant obstacles to research in this field. Furthermore, it is unknown whether the dynamic equations for thermal conduction and radiation satisfy transformation invariance. Therefore, the design of thermal metamaterials considering conduction and radiation remains an unexplored topic, even many years after the proposition of thermal metamaterials. Xu et al. were the first to design diffusion metamaterials for manipulating thermal radiation. For convenience, they did not use Stefan's law but instead employed the Rosseland diffusion approximation. Based on this method, they designed devices with novel functions such as cloaks, transparency, and expanders.

\begin{figure}[!ht]
  \includegraphics[width=1\linewidth]{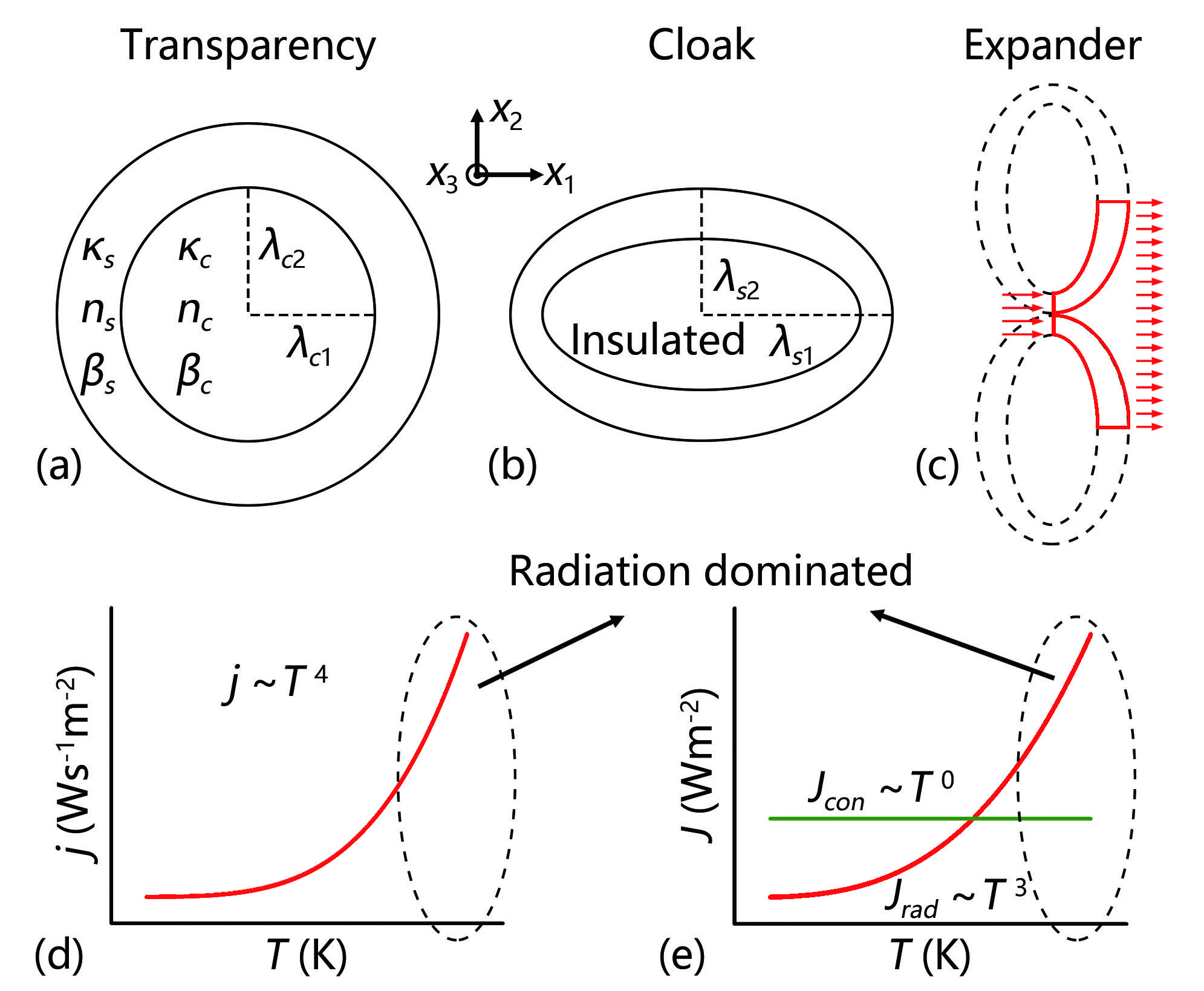}
  \caption{Schematic diagrams of (a) thermal transparency, (b) thermal cloak, and (c) thermal expander. (d) and (e) qualitatively show the radiative emittance $j$, conductive flux $J_{con}$, and radiative flux $J_{rad}$ as a function of temperature $T$. Adapted from Ref.~\cite{XuPRAP19}}
  \label{1}
  \end{figure}

 \section{Effective-medium theory under Rosseland approximation}
 Xu et al. investigated a passive and steady process of heat transfer, focusing on the total heat flux $\bm{J}_{\text{total}}$, which comprises the conductive flux $\bm{J}_{\text{con}}$ and the radiative flux $\bm{J}_{\text{rad}}$. This heat flux satisfies the divergence-free condition:

\begin{equation}\label{D}
\bm{\nabla}\cdot\bm{J}_{\text{total}} = \bm{\nabla}\cdot\left(\bm{J}_{\text{con}}+\bm{J}_{\text{rad}}\right) = 0.
\end{equation}

The conductive flux $\bm{J}_{\text{con}}$ is given by:

\begin{equation}\label{Jc}
\bm{J}_{\text{con}} = -\kappa\bm{\nabla}T,
\end{equation}

where $\kappa$ represents the thermal conductivity. On the other hand, based on the Rosseland diffusion approximation, the radiative flux $\bm{J}_{\text{rad}}$ is expressed as:

\begin{equation}\label{Jr}
\bm{J}_{\text{rad}} = -\gamma T^3\bm{\nabla}T,
\end{equation}

Here, $\gamma$ (given by $\gamma = 16\beta^{-1}n^2\sigma/3$) can be considered as the radiative coefficient. In this expression, $\beta$ corresponds to the Rosseland mean extinction coefficient, $n$ represents the relative refractive index, and $\sigma$ is the Stefan-Boltzmann constant ($\sigma = 5.67\times10^{-8}\,{\rm Wm}^{-2}{\rm K}^{-4}$).
   
Xu et al. further consider a three-dimensional core-shell structure (Fig.~\ref{1}(a)) which consists of a core with thermal conductivity $\kappa_c$, Rosseland mean extinction coefficient $\beta_c$, and relative refractive index $n_c$ (radiative coefficient $\gamma_c$), coated by a shell with corresponding parameters $\kappa_s$, $\beta_s$, and $n_s$ (radiative coefficient $\gamma_s$). The subscript $c$ (or $s$) denotes the core (or shell). The semi-axis lengths of the core and shell are $\lambda_{ci}$ and $\lambda_{si}$, respectively, where $i=1,\,2,\,3$. Assuming that the ratio $\gamma/\kappa$ of the core-shell structure is a constant $\alpha$, specifically $\gamma_c/\kappa_c=\gamma_s/\kappa_s=\alpha$, Equation~(\ref{D}) can be rewritten as

\begin{equation}\label{D1}
     \bm{\nabla}\cdot\left(-\kappa\bm{\nabla}T-\alpha\kappa T^3\bm{\nabla}T\right)=\bm{\nabla}\cdot\left(-\kappa\left(1+\alpha T^3\right)\bm{\nabla}T\right)=\bm{\nabla}\cdot\left(-\kappa\bm{\nabla}\left(T+\alpha T^4/4 \right)\right)=0.
\end{equation}

By performing a variable substitution $\varphi=T+\alpha T^4/4$, we obtain

\begin{equation}\label{D2}
     \bm{\nabla}\cdot\left(-\kappa\bm{\nabla}\varphi\right)=0.
\end{equation}

Therefore, the strongly nonlinear equation (Eq.~(\ref{D})) can be transformed into a linear equation (Eq.~(\ref{D2})).
   
To proceed, Xu et al. introduced ellipsoidal coordinates $\left(\rho,\,\xi,\,\eta\right)$, which are defined by the following equations:

\begin{equation}\label{ellipsoidal}
\left\{
\begin{array}{l}
\frac{x^2}{\rho+\lambda_1^2}+\frac{y^2}{\rho+\lambda_2^2}+\frac{z^2}{\rho+\lambda_3^2}=1\,\left({\rm confocal\ ellipsoids}\right)\\
\frac{x^2}{\xi+\lambda_1^2}+\frac{y^2}{\xi+\lambda_2^2}+\frac{z^2}{\xi+\lambda_3^2}=1\,\left({\rm hyperboloids\ of\ one\ sheet}\right)\\
\frac{x^2}{\eta+\lambda_1^2}+\frac{y^2}{\eta+\lambda_2^2}+\frac{z^2}{\eta+\lambda_3^2}=1\,\left({\rm hyperboloids\ of\ two\ sheets}\right)
\end{array}
\right.
\end{equation}

In these equations, $\lambda_1$, $\lambda_2$, and $\lambda_3$ are three constants that satisfy $\rho>-\lambda_1^2>\xi>-\lambda_2^2>\eta>-\lambda_3^2$. Equation~(\ref{D2}) can be expressed in ellipsoidal coordinates as:

\begin{equation}\label{Laplace}
\frac{\partial}{\partial\rho}\left(g\left(\rho\right)\frac{\partial \varphi}{\partial\rho}\right)+\frac{g\left(\rho\right)}{\rho+\lambda_i^2}\frac{\partial \varphi}{\partial\rho}=0,
\end{equation}

where $g\left(\rho\right)=\sqrt{\left(\rho+\lambda_1^2\right)\left(\rho+\lambda_2^2\right)\left(\rho+\lambda_3^2\right)}$.
Since Eq.~(\ref{Laplace}) has the solution:

\begin{equation}\label{solution}
\varphi=\left(u+v\int_0^\rho\left(\rho+\lambda_i^2\right)^{-1}g\left(\rho\right)^{-1}d\rho\right)x_i,
\end{equation}

where $u$ and $v$ are constants, and $x_i$ $\left(i=1,\,2,\,3\right)$ denotes Cartesian coordinates. The temperatures of the core, shell, and background can be defined as $\varphi_c$, $\varphi_s$, and $\varphi_b$, respectively. These temperatures can be obtained as follows:

\begin{equation}\label{temperature}
\left\{
\begin{array}{lll}
\varphi_c=u_c x_i \\
\varphi_s=\left(u_s+v_s\displaystyle\int_{\rho_c}^\rho\left(\rho+\lambda_i^2\right)^{-1}g\left(\rho\right)^{-1}d\rho\right)x_i\\
\varphi_b=u_b x_i
\end{array}
\right.
\end{equation}
where $u_c$, $u_s$, and $v_s$ are determined by the boundary conditions. The exterior surfaces of the core and shell are denoted by $\rho_c$ and $\rho_s$, respectively. The boundary conditions require continuity of temperatures and normal heat fluxes, leading to the following equations:

\begin{equation}\label{boundary}
\left\{
\begin{array}{llll}
u_c=u_s \\
u_b=u_s+v_s\displaystyle\int_{\rho_c}^{\rho_s}\left(\rho+\lambda_i^2\right)^{-1}g\left(\rho\right)^{-1}d\rho\\
u_c=2v_s\kappa_s\left(\kappa_c-\kappa_s\right)^{-1}g\left(\rho_c\right)^{-1}\\
u_b=2v_s\kappa_s\left(\kappa_{ei}-\kappa_s\right)^{-1}g\left(\rho_s\right)^{-1}
\end{array}
\right.
\end{equation}

Here, $\kappa_{ei}$ represents the effective thermal conductivity of the core-shell structure along the direction of $x_i$. To obtain the expression for $\kappa_{ei}$, one can solve Eq.~(\ref{boundary}). For this purpose, Xu et al. define the semi-axis lengths of the core, $\lambda_{ci}$, and the shell, $\lambda_{si}$, as follows:

\begin{equation}\label{length}
\left\{
\begin{array}{ll}
\lambda_{ci}=\sqrt{\lambda_i^2+\rho_c} \\
\lambda_{si}=\sqrt{\lambda_i^2+\rho_s}
\end{array}
\right.
\end{equation}

Here, $i=1,\,2,\,3$. Consequently, the volume fraction $f$ can be expressed as:

\begin{equation}\label{f}
f=\frac{\lambda_{c1}\lambda_{c2}\lambda_{c3}}{\lambda_{s1}\lambda_{s2}\lambda_{s3}}=\frac{g\left(\rho_c\right)}{g\left(\rho_s\right)}
\end{equation}
Xu et al. also introduce the shape factor $d_{wi}$ along the direction of $x_i$ as follows:

\begin{equation}\label{shape1}
d_{wi}=\frac{\lambda_{w1}\lambda_{w2}\lambda_{w3}}{2}\int_0^\infty\left(\tau+\lambda_{wi}^2\right)^{-1}\left(\left(\tau+\lambda_{w1}^2\right)\left(\tau+\lambda_{w2}^2\right)\left(\tau+\lambda_{w3}^2\right)\right)^{-1/2}d\tau,
\end{equation}

Here, the subscript $w$ can take the values $c$ or $s$, representing the shape factor of the core or shell, respectively. Subsequently, they can obtain the following expression:

\begin{equation}\label{de}
\begin{array}{cc}
\displaystyle\int_{\rho_c}^{\rho_s}\left(\rho+\lambda_i^2\right)^{-1}g\left(\rho\right)^{-1}d\rho=\int_{\rho_c}^\infty\left(\rho+\lambda_i^2\right)^{-1}g\left(\rho\right)^{-1}d\rho-\int_{\rho_s}^\infty\left(\rho+\lambda_i^2\right)^{-1}g\left(\rho\right)^{-1}d\rho\\
=2d_{ci}g\left(\rho_c\right)^{-1}-2d_{si}g\left(\rho_s\right)^{-1}.
\end{array}
\end{equation}

Finally, Xu et al. derive a concise expression for $\kappa_{ei}$ as follows:

\begin{equation}\label{kappa}
\kappa_{ei}=\kappa_s\left[\frac{f\left(\kappa_c-\kappa_s\right)}{\kappa_s+\left(d_{ci}-fd_{si}\right)\left(\kappa_c-\kappa_s\right)}+1\right].
\end{equation}

The method described above is the standard approach for calculating the Laplace equation. The shape factors satisfy the sum rule $d_{w1}+d_{w2}+d_{w3}=1$. In principle, the effective thermal conductivity of any core-shell structure can be obtained using Eq.~(\ref{kappa}) when the core-shell structure is confocal or concentric. In fact, Eq.~(\ref{kappa}) can be reduced to handle cylindrical (two-dimensional) cases by setting $\lambda_{w3}=\infty$, resulting in $d_{w1}=\lambda_{w2}/\left(\lambda_{w1}+\lambda_{w2}\right)$, $d_{w2}=\lambda_{w1}/\left(\lambda_{w1}+\lambda_{w2}\right)$, and $d_{w3}=0$ (the sum rule $d_{w1}+d_{w2}+d_{w3}=1$ is still satisfied).

Since $\gamma/\kappa$ is a constant, the effective radiative coefficient can be expressed as:

\begin{equation}\label{gamma}
\gamma_{ei}=\gamma_s\left[\frac{f\left(\gamma_c-\gamma_s\right)}{\gamma_s+\left(d_{ci}-fd_{si}\right)\left(\gamma_c-\gamma_s\right)}+1\right],
\end{equation}

Here, $\gamma_{ei}$ represents the effective radiative coefficient of the core-shell structure along the direction of $x_i$. By using Eqs.~(\ref{kappa}) and (\ref{gamma}), one can predict the effective thermal conductivity and effective radiative coefficient. However, in order to achieve the same effect of conduction and radiation, it is necessary to maintain $\gamma/\kappa$ as a constant.

\begin{figure}[!ht]
\includegraphics[width=1\linewidth]{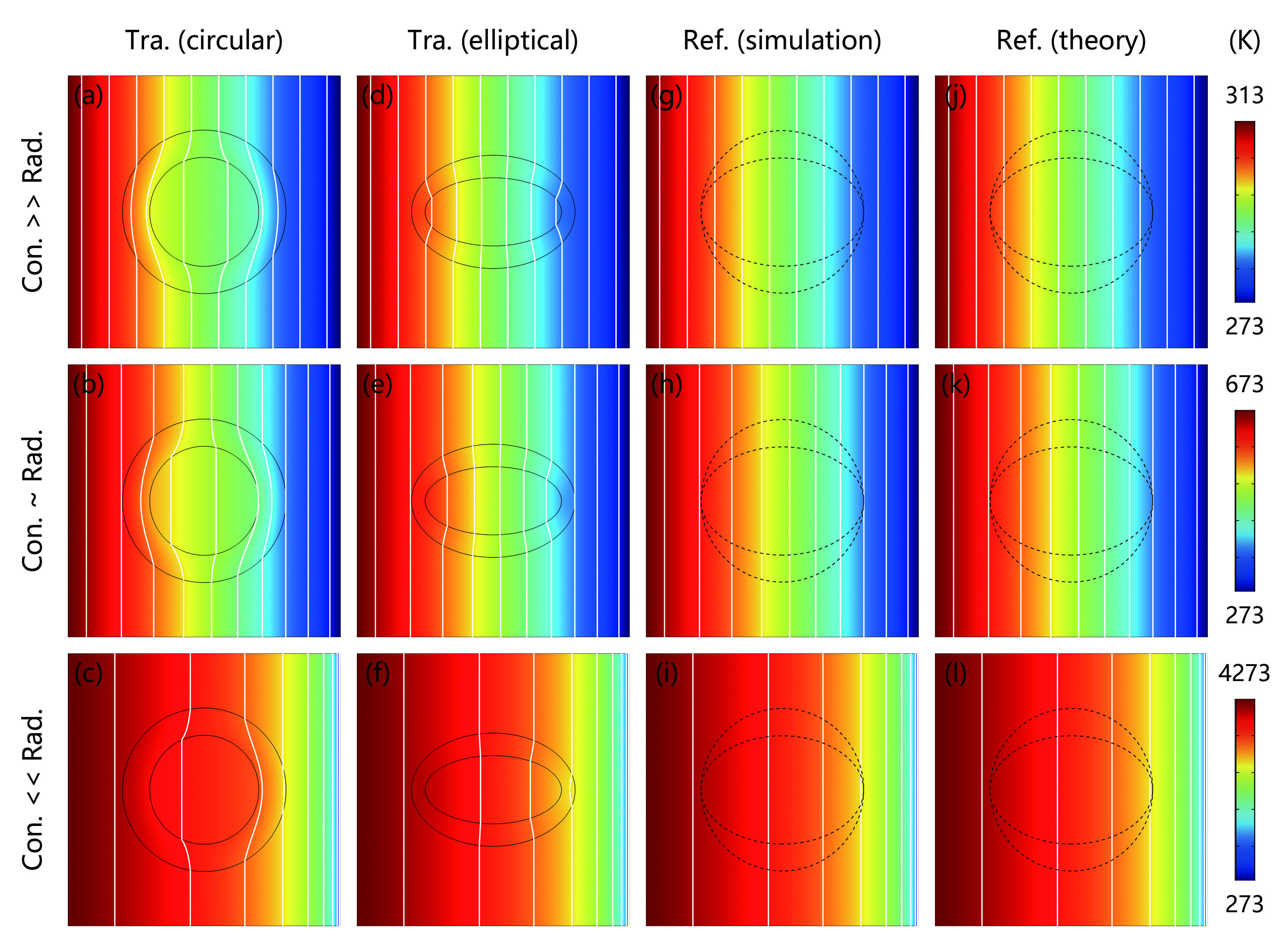}
\caption{Steady results of thermal transparency. (a)-(c): $\lambda_{c1}=\lambda_{c2}=2$~cm, $\lambda_{s1}=\lambda_{s2}=3$~cm, $\kappa_c=2$~Wm$^{-1}$K$^{-1}$, $\beta_c=50$~m$^{-1}$, $\kappa_s=0.62$~Wm$^{-1}$K$^{-1}$, and $\beta_s=161.1$~m$^{-1}$. (d)-(f): $\lambda_{c1}=2.5$~cm, $\lambda_{c2}=1.25$~cm, $\lambda_{s1}=3$~cm, $\lambda_{s2}=2.08$~cm, $\kappa_c=0.5$~Wm$^{-1}$K$^{-1}$, $\beta_c=200$~m$^{-1}$, $\kappa_s=1.61$~Wm$^{-1}$K$^{-1}$, and $\beta_s=62$~m$^{-1}$. (g)-(i) are the references with pure background parameters. (j)-(l) are the theoretical temperature distributions of the references. Circular (or elliptical) dashed lines are plotted for the comparison with circular (or elliptical) transparency. Adapted from Ref.~\cite{XuPRAP19}}
\label{2}
\end{figure}
Furthermore, to validate the theoretical analyses, Xu et al. conducted finite-element simulations. The radiative emittance $j$ is shown in Fig.~\ref{1}(d) to be proportional to $T^4$ according to the Stefan-Boltzmann law. The conductive flux $J_{\text{con}}$ depends on the temperature gradient, while the radiative flux $J_{\text{rad}}$ is proportional to $T^3$, as depicted in Fig.~\ref{1}(e). These qualitative analyses indicate the significant role of thermal radiation at high temperatures. Therefore, Xu et al. performed finite-element simulations at three temperature intervals: (I) 273-313 K, representing a small upper temperature limit where conduction (Con.) dominates; (II) 273-673 K, representing a medium upper temperature limit where conduction and radiation (Rad.) are approximately equal; and (III) 273-4273 K, representing a large upper temperature limit where radiation is the dominant mode of heat transfer.
\begin{figure}[!ht]
{\centering\includegraphics[width=0.8\linewidth]{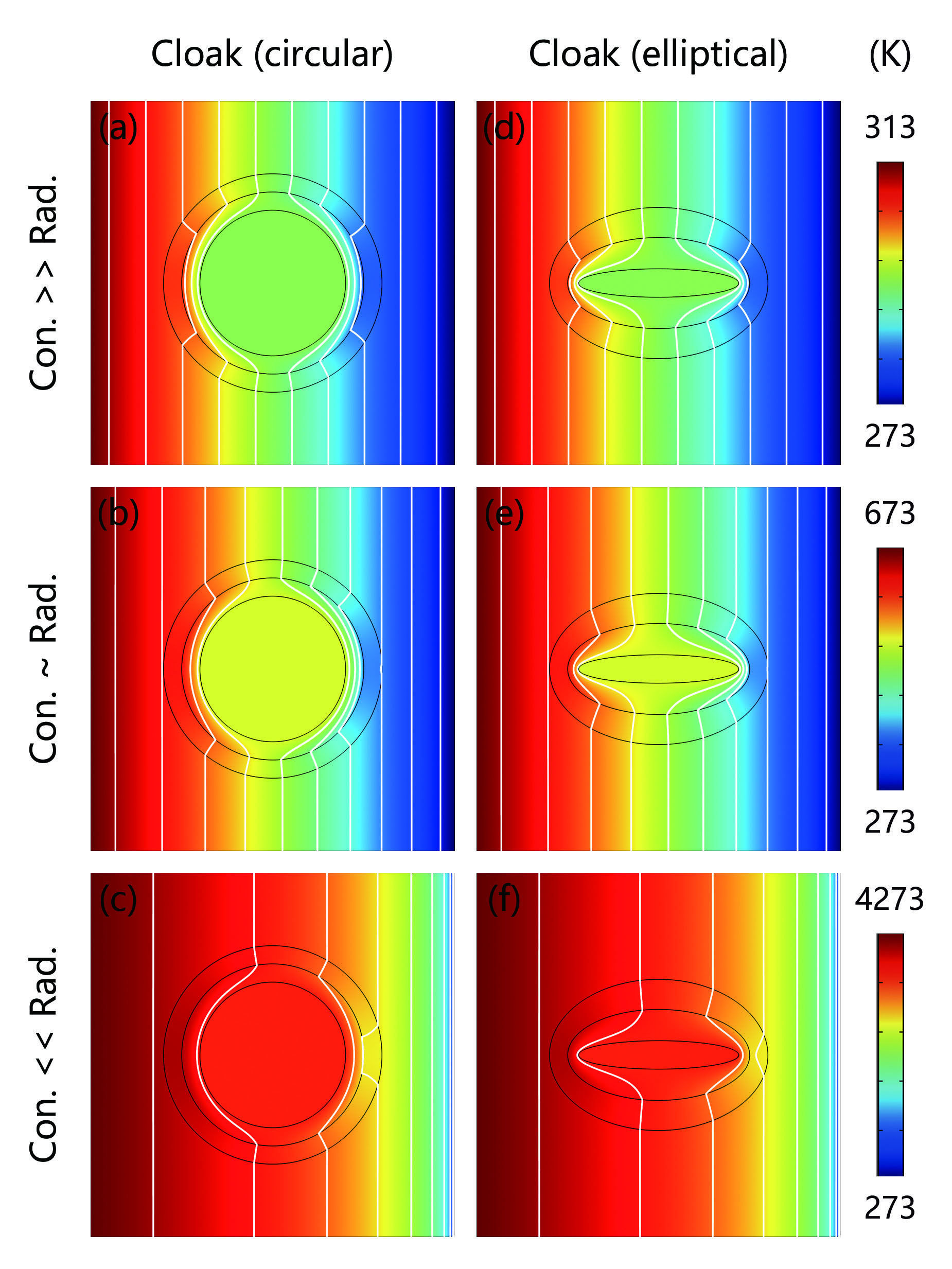}

}
\caption{Steady simulations of thermal cloak. An inner object is coated by an insulated layer with $\kappa=10^{-5}$~Wm$^{-1}$K$^{-1}$ and $\beta=10^5$~m$^{-1}$. Since the heat flux cannot enter into the insulated layer, the inner object plus the insulated layer can be equivalently regarded as an insulated core with $\kappa_c=10^{-5}$~Wm$^{-1}$K$^{-1}$ and $\beta_c=10^5$~m$^{-1}$. Other parameters are as follows. (a)-(c): $\lambda_{c1}=\lambda_{c2}=2.5$~cm, $\lambda_{s1}=\lambda_{s2}=3$~cm, $\kappa_s=5.54$~Wm$^{-1}$K$^{-1}$, and $\beta_s=18.1$~m$^{-1}$. (d)-(f): $\lambda_{c1}=2.5$~cm, $\lambda_{c2}=1.25$~cm, $\lambda_{s1}=3$~cm, $\lambda_{s2}=2.08$~cm, $\kappa_s=2.35$~Wm$^{-1}$K$^{-1}$, and $\beta_s=42.5$~m$^{-1}$. Adapted from Ref.~\cite{XuPRAP19}}
\label{3}
\end{figure}

Thermal transparency involves the design of a shell tailored to the object, ensuring an undistorted temperature profile outside the shell, as depicted in Fig.~\ref{2}. By selecting appropriate parameters based on Eqs.~(\ref{kappa}) and (\ref{gamma}), Xu et al. demonstrated that the temperature profile outside the shell remains undistorted, as shown in Figs.~\ref{2}(a)-\ref{2}(c) or Figs.~\ref{2}(d)-\ref{2}(f). Consequently, the core-shell structure at the center becomes indistinguishable, as illustrated in Figs.~\ref{2}(g)-\ref{2}(i). Figs.~\ref{2}(j)-\ref{2}(l) present the corresponding theoretical results from the references using the same parameters as Figs.~\ref{2}(g)-\ref{2}(i). The matching temperature profiles in both simulations and theory validate the theoretical analyses.

Furthermore, Xu et al. demonstrated that with a small upper temperature limit where conduction dominates, the temperature gradient outside the shell remains nearly uniform, as depicted in the first row of Fig.~\ref{2}. As the upper temperature limit increases, the effect of radiation becomes prominent, resulting in nonuniform temperature gradients outside the shell, as shown in the last two rows of Fig.~\ref{2}.

A thermal cloak is capable of shielding any object within it from detection. Generally, an insulated layer is employed to prevent heat flux from reaching the object. Consequently, the object along with the insulated layer can be treated as an insulated core, where $\kappa_c=\gamma=0$. Moreover, Xu et al. designed a shell based on Eqs.~(\ref{kappa}) and (\ref{gamma}) to eliminate the influence of the insulated core. The simulation results are presented in Figs.~\ref{3}(a)-\ref{3}(c) and Figs.~\ref{3}(d)-\ref{3}(f). Evidently, the isotherms remain separated from the object, indicating that the heat flux is unable to enter the object. Additionally, the temperature profiles of the background remain undistorted. This successful outcome demonstrates the expected cloaking effect.
\begin{figure}[!ht]
{\centering\includegraphics[width=0.8\linewidth]{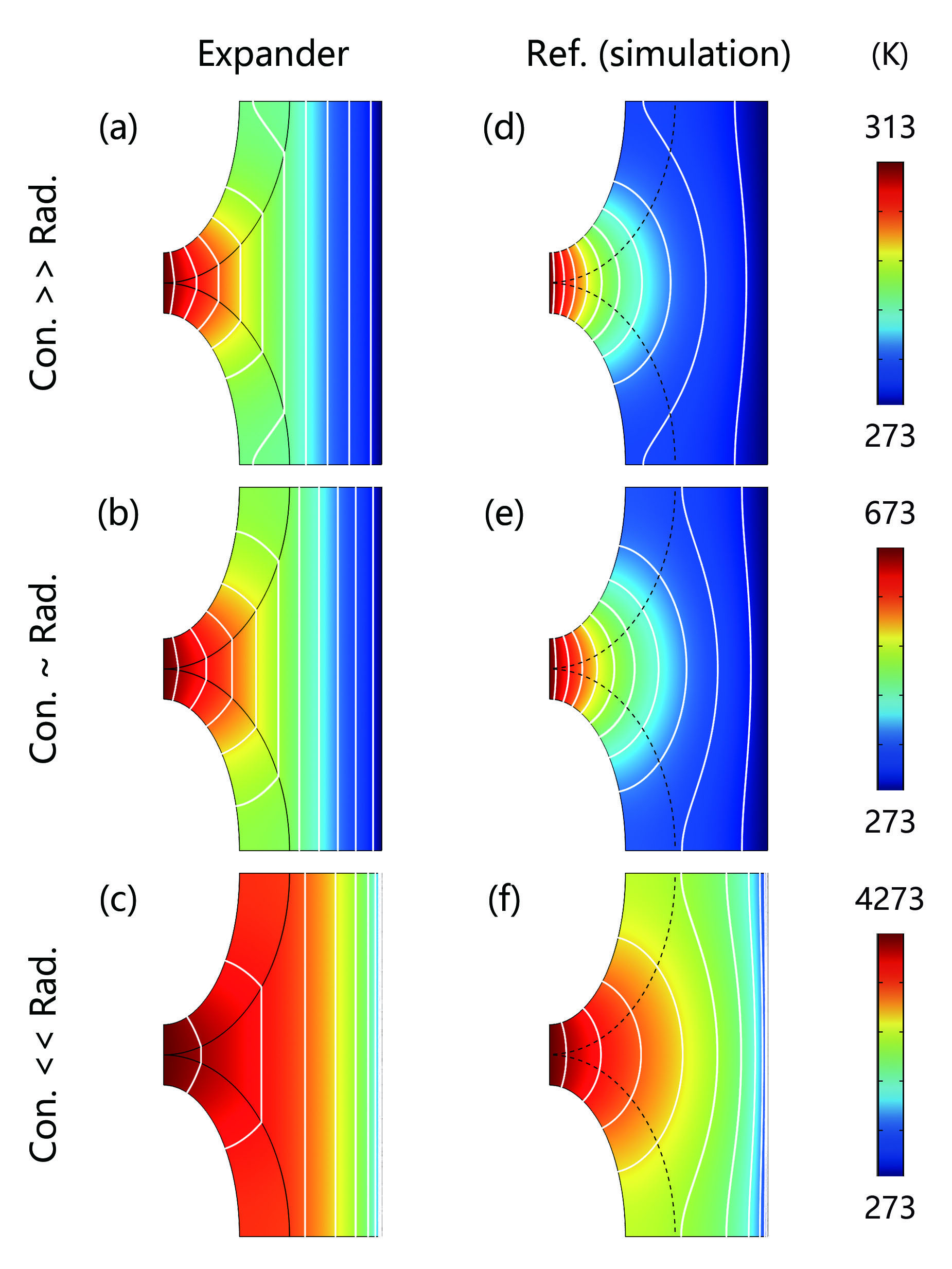}

}
\caption{Steady simulations of thermal expander. The sizes are $\lambda_{c1}=2.08$~cm, $\lambda_{c2}=4.17$~cm, $\lambda_{s1}=3.46$~cm, $\lambda_{s2}=5$~cm, and the width between hot and cold sources is 6~cm. Other parameters are as follows. (a)-(c): $\kappa_s=4.91$~Wm$^{-1}$K$^{-1}$ and $\beta_s=20.3$~m$^{-1}$. (d)-(f): pure background parameters. Adapted from Ref.~\cite{XuPRAP19}}
\label{4}
\end{figure}

The thermal expander concept aims to magnify a small heat source into a larger one by utilizing the design of two elliptical cloaks. To demonstrate this effect, Xu et al. assembled two elliptical cloaks together and extracted a quarter of the entire structure as the expander, as depicted in Fig.~\ref{1}(c). According to the uniqueness theorem in thermotics, the temperature distribution of the background remains undistorted, thereby achieving the desired expander effect. Finite-element simulations were conducted and presented in Figs.~\ref{4}(a)-\ref{4}(c). Clearly, the isotherms of the background appear as straight lines, indicating the excellent performance of the proposed structure. For comparison, Xu et al. also provided simulation results for a pure background material. These results reveal that the isotherms of the background become distorted, as shown in Figs.~\ref{4}(d)-\ref{4}(f).

\begin{figure}[!ht]
{\centering\includegraphics[width=0.8\linewidth]{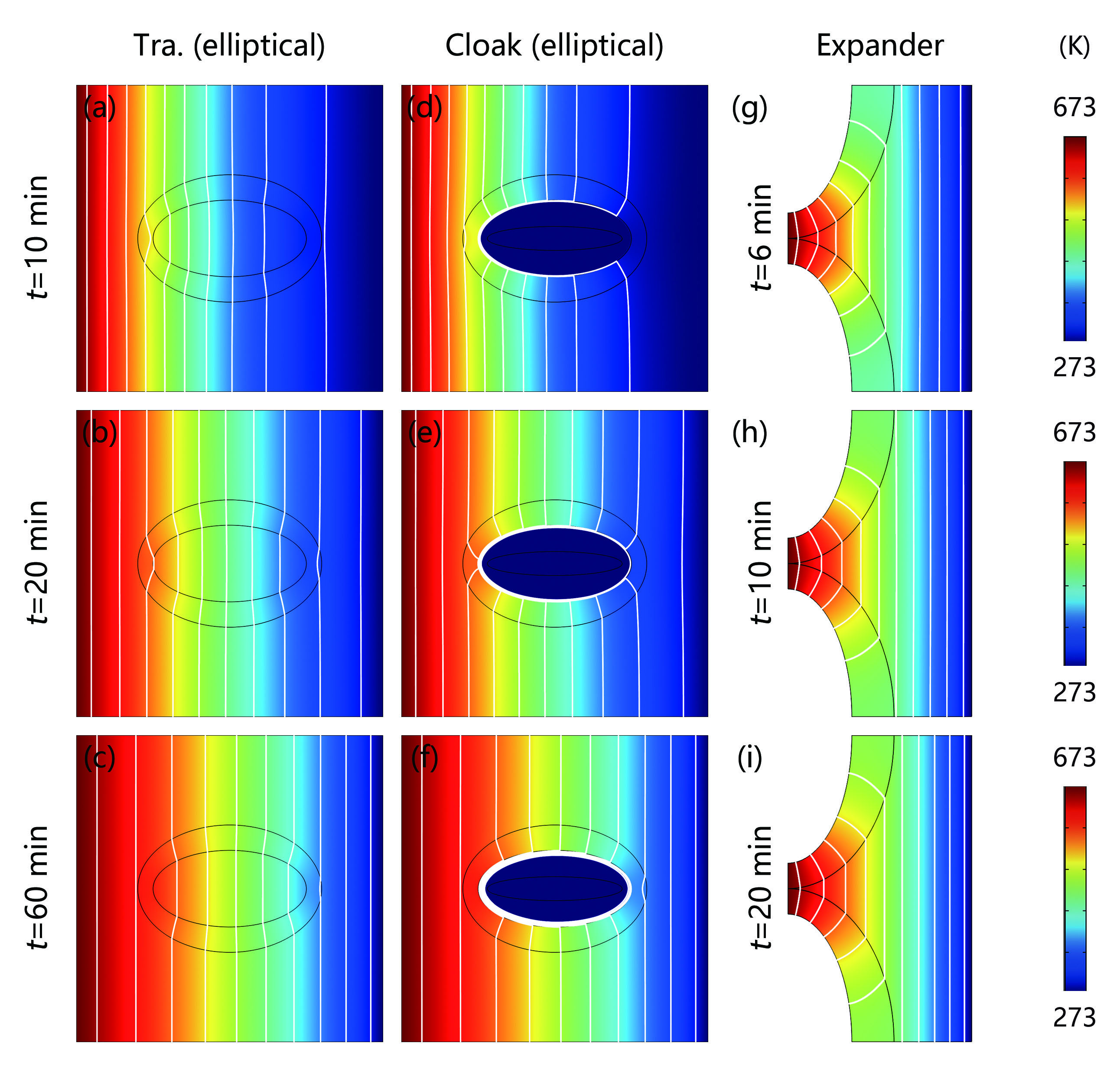}

}
\caption{Transient simulations of transparency, cloak, and expander. The sizes and material parameters of (a)-(c), (d)-(f), and (g)-(i) are the same as those for Figs.~\ref{2}(d)-\ref{2}(f), Figs.~\ref{3}(d)-\ref{3}(f), and Figs.~\ref{4}(a)-\ref{4}(c), respectively. The density and heat capacity of the background are $\rho C=10^6$~Jm$^{-3}$K$^{-1}$. Other parameters are as follows. (a)-(c): $\left(\rho C\right)_c=5\times10^5$~Jm$^{-3}$K$^{-1}$ and $\left(\rho C\right)_s=1.61\times10^6$~Jm$^{-3}$K$^{-1}$. (d)-(f): $\left(\rho C\right)_s=2.35\times10^6$~Jm$^{-3}$K$^{-1}$. (g)-(i): $\left(\rho C\right)_s=5\times10^5$~Jm$^{-3}$K$^{-1}$. Adapted from Ref.~\cite{XuPRAP19}}
\label{5}
\end{figure}

Xu et al. demonstrated that the theoretical analyses are applicable not only to steady states but also to transient cases. To illustrate this point, Xu et al. considered density and heat capacity. In order to design transient transparency and cloak, the value of heat diffusivity $\kappa/\left(\rho C\right)$ was set as a constant. Based on the simulation results, the performance of this approach remained satisfactory. The results at $t=10,\,20,\,60$~mins are depicted in Figs.~\ref{5}(a)-\ref{5}(c) and Figs.~\ref{5}(d)-\ref{5}(f), respectively.

To showcase the effect of transient expansion and achieve the optimal transient effect, Xu et al. employed an optimization method and set the diffusivity of the shell to be larger than that of the background. The results at $t=6,\,10,\,20$~mins are presented in Figs.~\ref{5}(g)-\ref{5}(i).

\section{Potential applications of radiative metamaterials: thermal camouflage and radiative cooler}
The development of diffusion metamaterials for radiation control holds significant importance for our future lives, owing to its potential applications such as thermal camouflage~\cite{Tan27,Tan50,Tan51,Tan52,Tan53,XuPLA18,XuJAP18,WangIJTS18} and radiative cooling. Thermal camouflage refers to a device that not only prevents the detection of objects but also generates deceptive signals. Researchers have designed systems that can mislead the detection of cloaked objects. Additionally, encrypted thermal printing has been achieved, enabling object detection only when an appropriate heat source is applied. While the original structure for thermal camouflage is two-dimensional, there are also studies focusing on realizing the same functionality using three-dimensional structures.

Moreover, radiative cooling is another application field worth mentioning. Radiative cooling involves the automatic achievement of cooling effects without the need for an external source. The concept was first proposed in 1978, but at that time, it remained primarily theoretical. It was in 2014 that a practical radiative cooler was designed based on the photonic approach. The device operates by selectively reflecting electromagnetic waves in the mid-infrared range. Since then, researchers have developed more efficient methods to achieve the same functionality. Experimental results have also confirmed the effectiveness of these devices. We believe that in the future, radiative coolers could find widespread use in our daily lives, provided that the costs are reduced to a certain level.

\section{Outlook: radiative metamaterials from microscopic view}
Over the past decades, research on diffusion metamaterials has undergone significant changes. Initially, the focus was on single-function devices, but it has since shifted towards multi-functionality~\cite{ZhuangIJMSD23,Tan15}. Similarly, there has been a transition from linear to nonlinear response~\cite{ZhuangPRE22,DaiJNU21,SuEPL20,DaiIJHMT20,YangPRE19,DaiEPJB18}, from temperature-independent to temperature-dependent devices~\cite{Tan16,LeiEPL21,Tan11}, from single thermal to thermoelectrical effects~\cite{QuEPL21,XuEPJB20,WangPRA19,Tan10}, from spatial to spatiotemporal metamaterials~\cite{YangPRA23,LeiMTP23,YangPRAP22,XuPRL221,XuPRE21}, and from temperature diffusion to temperature waves~\cite{ZhangTSEP21,XuEPL212,XuCPL20,XuAPL20,XuIJHMT20}. Furthermore, the research methodology has been applied to other fields as well, including hydrodynamics~\cite{LiPF22,WangCPB22,DaiPRAP22,WangPRAP21,WangATE21,WangICHMT20,DaiPRE18}, plasma physics~\cite{ZhangCPL22}, mass diffusion~\cite{ZhangPRA23,ZhangATS22}, and topology~\cite{JinPNAS23}. Novel concepts such as programmable metamaterials~\cite{LeiIJHMT23,ZhouESEE19}, nonreciprocity metamaterials~\cite{XuAPL21}, intelligent metamaterials~\cite{XuSCPMA20,YangEPL191,YangEPL192,XuPRA19,XuEPL19,XuEPJB192,XuPRE18,Tan4}, dipole-assisted thermotics~\cite{XuPRE191}, and negative thermal transport~\cite{XuCPLEL20} have also been proposed. Additionally, the research on networks~\cite{ShangIJHMT18,Tan2} and phase transitions~\cite{ZhouEPL23,LinSCPMA22,GaoNM21,HuangFP17} is expected to make further progress. In the future, the impact of diffusion metamaterials on the development of thermal diodes~\cite{Tan40,Tan41,Tan42,ShangJHT18,XuEPJB17} is eagerly anticipated. Machine learning~\cite{ZhangPRD22,LiuJAP212} is also increasingly being incorporated into the research methodology. While significant progress has been made in the study of diffusion metamaterials for conduction and radiation, the research has primarily focused on macroscopic scales. Exploring the same effects at the microscopic scale is an intriguing topic~\cite{Tan44}. To validate the Fourier law from a microscopic perspective, researchers have designed various models. The simplest among them is the harmonic lattice model~\cite{Tan55,Tan56,Tan57,Tan58,Tan59,Tan60,Tan61,Tan62,Tan63,Tan64,Tan65,Tan66,Tan67,Tan68}, which includes one-dimensional ordered cases~\cite{Tan69}, higher-dimensional ordered cases, one-dimensional disordered cases~\cite{Tan70,Tan71,Tan72,Tan73,Tan74,Tan75,Tan76,Tan77,Tan78,Tan79}, and two-dimensional disordered cases~\cite{Tan80,Tan81,Tan82,Tan83,Tan84,Tan85,Tan86,Tan122}. Additionally, the harmonic lattices with self-consistent reservoirs have been studied~\cite{Tan87,Tan88,Tan89,Tan90,Tan91,Tan92}. More complex models involve interacting systems~\cite{Tan93,Tan94,Tan95,Tan96,Tan97,Tan98}. The research in this field can be divided into two parts: momentum-conserving models, such as the FPU model~\cite{Tan99,Tan100,Tan101,Tan102,Tan103,Tan104,Tan106}, and momentum-non-conserving models~\cite{Tan107,Tan108,Tan109,Tan110,Tan111,Tan112,Tan113,Tan114,Tan115,Tan116,Tan117,Tan118,Tan119,Tan120,Tan121}. Some of these models provide essential insights into illustrating the Fourier law. Therefore, exploring the possibility of achieving novel functions such as cloaking and concentration in a similar manner is a worthwhile topic to explore. Furthermore, the experimental validation of microscopic theories poses a significant challenge. However, advancements in nanotechnology inspire us to move forward, and the combination of diffusion metamaterials with low-dimensional systems offers a new research field.

% \section{section2}\label{dsfdsfsf}

\bibliographystyle{model1-num-names}
% \bibliographystyle{cas-model2-names}

% \bibliography{sensor}
%Control: production of article title (0) alloXu et al.d
\balance

\end{document}